\documentclass[12pt]{article}
\usepackage{epsfig}
\usepackage{amsmath}
\usepackage{latexsym}
\title{Quantum Corrections to the Schwarzschild and
Kerr Metrics: Spin 1}
\author{Barry R. Holstein\\
Department of Physics-LGRT\\
University of Massachusetts\\
Amherst, MA  01003}
\begin{document}
\begin{titlepage}
\maketitle
\begin{abstract}
A previous evaluation of one-graviton loop corrections to the
energy-momentum tensor has been extended to particles with unit
spin and speculations are presented concerning general properties
of such forms.
\end{abstract}

\end{titlepage}

\section{Introduction}

In an earlier paper, we described calculations of the
graviton-loop corrections to the energy-momentum tensor of a
charged spinless or a spin 1/2 particle of mass $m$ and we focused
on the nonanalytic component of such results\cite{gar}.   This is
because such nonanalytic pieces involve singularities at small
momentum transfer $q$ which, when Fourier-transformed, yield---via
the Einstein equations---large distance corrections to the metric
tensor. In particular, for both a spinless field and for a spin
1/2 field the diagonal components of the metric were shown to be
modified from their simple Schwarzschild or Kerr forms---in
harmonic gauge
\begin{eqnarray}
g_{00}&=&1-{2Gm\over r}+{2G^2m^2\over r^2}+{7G^2m\hbar\over \pi
r^3}+\ldots\nonumber\\
g_{ij}&=&-\delta_{ij}[1+{2Gm\over r}+{G^2m^2\over r^2}
+{14G^2m\hbar\over 15\pi r^3}-{76\over 15}{G^2m\hbar\over \pi r^3
}(1-\log\mu r)]\nonumber\\
&-&{r_ir_j\over r^2}[{G^2m^2\over r^2}+{76G^2m\hbar\over 15\pi
r^3} +{76\over 5}{G^2m\hbar\over \pi r^3}(1-\log\mu r)]
\end{eqnarray}
where $G$ is the gravitational constant.   (Note that the
dependence on the arbitrary scale factor $\mu$ can be removed by a
coordinate transformation.) The
classical---$\hbar$-independent---pieces of these modifications
are well known and can be found by expanding the familiar
Schwarzschild (Kerr) metric, which describes spacetime around a
massive (spinning) object\cite{rnm}. On the other hand, the
calculation also yields quantum
mechanical---$\hbar$-dependent---pieces which are new and whose
origin can be understood qualitatively as arising from
zitterbewegung\cite{gar}.

\begin{figure}[h]
\begin{center}
\epsfig{file=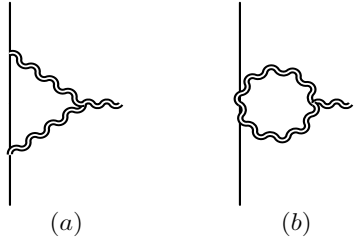,height=4cm,width=7cm} \caption{Feynman
diagrams having nonanalytic components.  Here
 the doubly wiggly lines represent gravitons.}
\end{center}
\end{figure}

In the case of a spin 1/2 system there exists, in addition to the
above, a nonvanishing {\it off}-diagonal piece of the metric,
whose one-loop corrected form, in harmonic gauge, was found to be
\begin{equation}
g_{ij}=(\vec{S}\times\vec{r})_i\left({2G\over r^3}-{2G^2m\over
r^4}+{3G^2\hbar\over \pi r^5} +\ldots\right)
\end{equation}
Here the classical component of this modification can be found by
expanding the Kerr metric\cite{knm}, describing spacetime around a
spinning mass and once again there exist quantum corrections due
to zitterbewegung\cite{gar}.

Based on the feature that the diagonal components were found to
have an identical form for both spin 0 and 1/2, it is tempting to
speculate that the leading diagonal piece of the metric about a
charged particle has a universal form---independent of spin.
Whether the same is true for the leading
off-diagonal---spin-dependent---component cannot be determined
from a single calculation, but it is reasonable to speculate that
this is also the case. However, whether these assertions are
generally valid can be found only by further calculation, which is
the purpose of the present note, wherein we evaluate the
nonanalytic piece of the graviton-loop-corrected energy-momentum
tensor for a particle of spin 1 and assess the correctness of our
proposal.  In the next section then we briefly review the results
of the previous paper, followed by a discussion wherein the
calculations are extended to the spin 1 system.  Results are
summarized in a brief concluding section.

\section{Lightning Review}

Since it important to the remainder of this note, we first present
a brief review of the results  obtained in our previous
paper\cite{gar}.  In the case of spin 0 systems, the general form
of the energy-momentum tensor is
\begin{equation}
<p_2|T_{\mu\nu}(x)|p_1>_{S=0}={e^{i(p_2-p_1)\cdot x}\over
\sqrt{4E_2E_1}}\left[2P_\mu P_\nu F_1^{(S=0)}(q^2))+(q_\mu
q_\nu-q^2\eta_{\mu\nu})F_2^{(S=0)}(q^2)\right]
\end{equation}
where $P={1\over 2}(p_1+p_2)$ is the average momentum while
$q=p_1-p_2$ is the momentum transfer.  The tree level values for
these form factors are
\begin{equation}
F_{1,tree}^{(S=0)}=1\qquad F_{2,tree}^{(S=0)}=-{1\over 2}
\end{equation}
while the leading nonanalytic loop corrections from Figure 1a and
Figure 1b were determined to be
\begin{eqnarray}
F_{1,loop}^{(S=0)}(q^2)&=& {Gq^2\over \pi}\left(-{3\over 4}L
+{1\over 16}S\right) \nonumber\\
F_{2,loop}^{(S=0)}(q^2)&=&{Gm^2\over \pi}\left(-2L+{7\over
8}S\right)
\end{eqnarray}
where we have defined
$$L=\log ({-q^2\over m^2})\quad {\rm and}\quad S=\pi^2\sqrt{m^2\over
-q^2}.$$ Such pieces, which are singular in the small-q limit,
come about due to the presence of two massless propagators in the
Feynman diagrams\cite{doh} and can arise even in electromagnetic
diagrams when this situation is present\cite{emf}.  Upon
Fourier-transforming, the component proportional to $S$ is found
to yield classical ($\hbar$-independent) behavior while the term
involving $L$ yields quantum mechanical ($\hbar$-dependent)
corrections. The feature that the form factor $F_1^{(S=0)}(q^2=0)$
remains unity even when graviton loop corrections are included
arises from the stricture of energy-momentum
conservation\cite{gar}.  There exists no restriction on
$F_2^{(S=0)}(q^2=0)$.

In the case of spin 1/2 there exists an additional form
factor---$F_3^{(S={1\over 2})}(q^2)$---associated with the
presence of spin---
\begin{eqnarray}
<p_2|T_{\mu\nu}(x)|p_1>_{S={1\over 2}}&=&{e^{i(p_2-p_1)\cdot
x}\over \sqrt{E_1E_2}} \bar{u}(p_2)\left[P_\mu P_\nu
F_1^{(S={1\over
2})}(q^2)\right.\nonumber\\
&+&\left.{1\over 2}(q_\mu q_\nu -q^2\eta_{\mu\nu})F_2^{(S={1\over
2
)}}(q^2)\right.\nonumber\\
&-&\left.\left({i\over 4}\sigma_{\mu\lambda}q^\lambda
P_\nu+{i\over 4} \sigma_{\nu\lambda}q^\lambda
P_\mu\right)F_3^{(S={1\over 2})}(q^2)\right]u(p_1)
\end{eqnarray}
In this case, the tree level values for these form factors are
\begin{equation}
F_{1,tree}^{(S={1\over 2})}=F_{2,tree}^{(S={1\over 2})}=1\qquad
F_{3,tree}^{(S={1\over 2})}=0
\end{equation}
while the nonanalytic loop corrections from Figure 1a and Figure
1b were determined to be
\begin{eqnarray}
F_{1,loop}^{(S={1\over 2})}(q^2)&=&{Gq^2\over \pi}(-{3\over 4}L
+{1\over 16}S) \nonumber\\
F_{2,loop}^{(S={1\over 2})}(q^2)&=&{Gm^2\over \pi}(-2L+{7\over 8}S)\nonumber\\
F_{3,loop}^{(S={1\over 2})}(q^2)&=&{Gq^2\over \pi}({1\over
4}L+{1\over 4}S)
\end{eqnarray}
In this case both $F_1^{(S={1\over 2})}(q^2=0)$ {\it and}
$F_3^{(S={1\over 2})}(q^2=0)$ retain their value of unity even in
the presence of graviton loop corrections.  That this must be true
for $F_1^{(S={1\over 2})}(q^2=0)$ follows from energy-momentum
conservation, as before, while the nonrenormalization of
$F_3^{(S={1\over 2})}(q^2=0)$ is required by angular-momentum
conservation\cite{gar}.  An interesting consequence is that there
{\it cannot} exist an anomalous gravitomagnetic moment.  The
universality of these radiative corrections is suggested by the
results
\begin{equation}
F_{1,loop}^{(S=0)}(q^2)=F_{1,loop}^{(S={1\over 2})}(q^2)\quad{\rm
and} \quad F_{2,loop}^{(S=0)}(q^2)=F_{2,loop}^{(S={1\over
2})}(q^2)
\end{equation}
but, of course, the spin-dependent gravitomagnetic form factor
$F_3^{(S={1\over 2})}(q^2)$ has no analog in the spin 0 sector.

The connection with the metric tensor described in the
introduction arises when these results for the energy-momentum
tensor are combined with the (linearized) Einstein
equation\cite{eeq}
\begin{equation}
\Box h_{\mu\nu}=-16\pi G\left(T_{\mu\nu}-{1\over 2} \eta_{\mu\nu}
T\right)
\end{equation}
where we have defined
\begin{equation}
g_{\mu\nu}=\eta_{\mu\nu}+h_{\mu\nu}
\end{equation}
and
\begin{equation}
T\equiv {\rm Tr}\,T_{\mu\nu}
\end{equation}
Taking Fourier transforms, we
find---for both spin 0 and spin 1/2---the diagonal
components\footnote{Here the r-dependent corrections proportional
to $\hbar$ arise from the graviton vacuum polarization correction,
while those independent of $\hbar$ arise from corrections to the
linear Einstein equation\cite{gar}.}
\begin{eqnarray}
h_{00}(\vec{r})&=&-16\pi G\int{d^3k\over
(2\pi)^3}e^{i\vec{k}\cdot\vec{r}}{1\over \vec{k}^2}\left({m\over
2}- {Gm^2\pi|\vec{k}|\over 4}+{7Gm\vec{k}^2\over
8\pi}\log{\vec{k}^2\over m^2}\right)
-{43G^2m\hbar\over 15\pi r^3}\nonumber\\
h_{ij}(\vec{r})&=&-16\pi G\int{d^3k\over
(2\pi)^3}e^{i\vec{k}\cdot\vec{r}}{1\over \vec{k}^2}\left[{m\over
2} \delta_{ij}-\delta_{ij}\left({Gm^2\pi|\vec{k}|\over 32}
-{3Gm\vec{k}^2\over 8\pi}\log{\vec{k}^2\over m^2}\right)\right.\nonumber\\
&+&\left.\left(k_ik_j+{1\over
2}\vec{k}^2\delta_{ij}\right)\left({7Gm^2\pi\over
16|\vec{k}|}-{Gm\over \pi}\log\vec{k}^2\right)\right]\nonumber\\
&+&4G^2m^2\left({\delta_{ij}\over r^2}-2{r_ir_j\over r^4}\right)
+{G^2m\hbar\over 15\pi r^3}(\delta_{ij}+44{r_ir_j\over
r^2})\nonumber\\
&-& {44G^2m\hbar\over 15\pi r^3}(\delta_{ij}-3{r_ir_j\over
r^2})(1-\log\mu r)
\end{eqnarray}
while in the case of the spin 1/2 gravitomagnetic form factor we
find the off-diagonal term
\begin{eqnarray}
h_{0i}(\vec{r})&=&-16\pi G{i\over 2}\int{d^3k\over
(2\pi)^3}{1\over \vec{k}^2} \left(1-{Gm\pi|\vec{k}|\over
4}-{G\vec{k}^2\over 4\pi}
\log{\vec{k}^2\over m^2}\right)(\vec{S}\times\vec{k})_i\nonumber\\
&+&{21G^2\hbar\over 5\pi r^5}(\vec{S}\times\vec{r})_i
\end{eqnarray}
Evaluating the various Fourier transforms, we find the results
quoted in the introduction\cite{fta}.

The purpose of the present note is to study how these results
generalize to the case of higher spin.  Specifically, we shall
below examine the graviton-loop corrections to the energy-momentum
tensor of a massive spin 1 system.

\section{Spin 1}

A neutral spin 1 field $\phi_\mu(x)$ having mass $m$ is described
by the Proca Lagrangian\cite{pro}
\begin{equation}
{\cal L}(x)=-{1\over 4}U_{\mu\nu}(x)U^{\mu\nu}(x)+{1\over
2}m^2\phi_\mu(x)\phi^\mu(x)\label{eqn:la}
\end{equation}
where
\begin{equation}
U_{\mu\nu}(x)=i\partial_\mu \phi_\nu(x)-i\partial_\nu\phi_\mu(x)
\end{equation}
is the spin 1 field tensor.  Having the Langrangian for the
interactions of a spin-1 system, we can calculate the matrix
elements which will be required for our calculation. Specifically,
the general single graviton vertex for a transition involving an
outgoing graviton with polarization indices $\mu\nu$ and
four-momentum $q=p_1-p_2$, an incoming spin one particle with
polarization index $\alpha$ and four-momentum $p_1$ together with
an outgoing spin one particle with polarization index $\beta$ and
four-momentum $p_2$ is
\begin{eqnarray}
V^{(1)}_{\beta,\alpha,\mu\nu}(p_1,p_2)&=&i{\kappa\over
2}\left\{(p_{1\mu}p_{2\nu}+p_{1\nu}p_{2\mu})\eta_{\alpha\beta}+\eta_{\mu\nu}p_{1\beta}p_{2\alpha}
\right.\nonumber\\
&-&\left.p_{1\beta}(p_{2\mu}\eta_{\nu\alpha}+p_{2\nu}\eta_{\alpha\mu})
-p_{2\alpha}(p_{1\mu}\eta_{\nu\beta}+p_{1\nu}\eta_{\beta\mu})\right.\nonumber\\
&+&\left.(p_1\cdot p_2-m^2)(\eta_{\mu\alpha}\eta_{\nu\beta}+
\eta_{\mu\beta}\eta_{\nu\alpha}-\eta_{\mu\nu}\eta_{\alpha\beta})\right\}\label{eqn:to}
\end{eqnarray}
where $\kappa=\sqrt{32\pi G}$ is the gravitational coupling, while
the two-graviton vertex with polarization indices $\mu\nu$ and
$\rho\sigma$, an incoming spin one particle with polarization
index $\alpha$ and four-momentum $p_1$ together with an outgoing
spin one particle with polarization index $\beta$ and
four-momentum $p_2$ has the form
\begin{eqnarray}
V^{(2)}_{\beta,\alpha,\mu\nu,\rho\sigma}(p_1,p_2)&=&-i{\kappa^2\over
4}\left\{[p_{1\beta}p_{2\alpha} -
          \eta_{\alpha\beta}(p_1\cdot p_2 - m^2)]
      (\eta_{\mu\rho}\eta_{\nu\sigma}+
          \eta_{\mu\sigma}\eta_{\nu\rho} -
          \eta_{\mu\nu}\eta_{\rho\sigma})\right.\nonumber\\
          &+&\left.
    \eta_{\mu\rho}[\eta_{\alpha\beta}(p_{1\nu}p_{2\sigma} +
                p_{1\sigma}p_{2\nu}) -
          \eta_{\alpha\nu}p_{1\beta}p_{2\sigma}-
          \eta_{\beta\nu}p_{1\sigma}p_{2\alpha}\right.\nonumber\\
          &-&\left.
          \eta_{\beta\sigma}p_{1\nu}p_{2\alpha} -
          \eta_{\alpha\sigma}p_{1\beta}
            p_{2\nu} + (p_1\cdot p_2 -
                m^2)(\eta_{\alpha\nu}\eta_{\beta\sigma} +
                \eta_{\alpha\sigma}\eta_{\beta\nu})]\right.\nonumber\\
                &+&\left.
    \eta_{\mu\sigma}[\eta_{\alpha\beta}(p_{1\nu}p_{2\rho} +
                p_{1\rho}p_{2\nu}) -
          \eta_{\alpha\nu}p_{1\beta}p_{2\rho} -
          \eta_{\beta\nu}p_{1\rho}p_{2\alpha}\right.\nonumber\\
          &-&\left.
          \eta_{\beta\rho}p_{1\nu}p_{2\alpha}-
          \eta_{\alpha\rho}p_{1\beta}
            p_{2\nu} + (p_1\cdot p_2 -
                m^2)\eta_{\alpha\nu}\eta_{\beta\rho} +
                \eta_{\alpha\rho}\eta_{\beta\nu})]\right.\nonumber\\
                &+&\left.
    \eta_{\nu\rho}[\eta_{\alpha\beta}(p_{1\mu}p_{2\sigma} +
                p_{1\sigma}p_{2\mu})
                -\eta_{\alpha\mu}p_{1\beta}p_{2\sigma} -
          \eta_{\beta\mu}p_{1\sigma}p_{2\alpha}\right.\nonumber\\
          &-&\left.\eta_{\beta\sigma}p_{1\mu}p_{2\alpha}
          -\eta_{\alpha\sigma}p_{1\beta}
            p_{2\mu} + (p_1\cdot p_2 -
                m^2)(\eta_{\alpha\mu}\eta_{\beta\sigma} +
                \eta_{\alpha\sigma}\eta_{\beta\mu})]\right.\nonumber\\
                &+&\left.
    \eta_{\nu\sigma}[\eta_{\alpha\beta}(p_{1\mu}p_{2\rho} +
                p_{1\rho}p_{2\mu}) -
          \eta_{\alpha\mu}p_{1\beta}p_{2\rho} -
          \eta_{\beta\mu}p_{1\rho}p_{2\alpha}\right.\nonumber\\
          &-&\left.\eta_{\beta\rho}p_{1\mu}p_{2\alpha}-\eta_{\alpha\rho}p_{1\beta}
            p_{2\mu} + (p_1\cdot p_2 -
                m^2)(\eta_{\alpha\mu}\eta_{\beta\rho} +
                \eta_{\alpha\rho}\eta_{\beta\mu})]\right.\nonumber\\
                &-&\left.
    \eta_{\mu\nu}[\eta_{\alpha\beta}(p_{1\rho}p_{2\sigma} +
                p_{1\sigma}p_{2\rho}) -
          \eta_{\alpha\rho}p_{1\beta}p_{2\sigma} -
          \eta_{\beta\rho}p_{1\sigma}p_{2\alpha}\right.\nonumber\\
          &-&\left.\eta_{\beta\sigma}p_{1\rho}p_{2\alpha}-
          \eta_{\alpha\sigma}p_{1\beta}p_{2\rho} + (p_1\cdot p_2 -
                m^2)(\eta_{\alpha\rho}\eta_{\beta\sigma} +
                \eta_{\beta\rho}\eta_{\alpha\sigma})]\right.\nonumber\\
                &-&\left.
    \eta_{\rho\sigma}[\eta_{\alpha\beta}(p_{1\mu}p_{2\nu} +
                p_{1\nu}p_{2\mu}) -
          \eta_{\alpha\mu}p_{1\beta}p_{2\nu} -
          \eta_{\beta\mu}p_{1\nu}p_{2\alpha}\right.\nonumber\\
          &-&\left.
          \eta_{\beta\nu}p_{1\mu}p_{2\alpha} -
          \eta_{\alpha\nu}p_{1\beta}
            p_{2\mu} + (p_1\cdot p_2 -
                m^2)(\eta_{\alpha\mu}\eta_{\beta\nu} +
                \eta_{\beta\mu}\eta_{\alpha\nu})]\right.\nonumber\\
                 &+&\left.
    (\eta_{\alpha\rho}p_{1\mu} -
          \eta_{\alpha\mu}p_{1\rho})(\eta_{\beta\sigma}
            p_{2\nu} - \eta_{\beta\mu}p_{2\sigma})\right.\nonumber\\
            &+&\left.
    (\eta_{\alpha\sigma}p_{1\nu} -
          \eta_{\alpha\nu}p_{1\sigma})\eta_{\beta\rho}
            p_{2\mu} - \eta_{\beta\mu}p_{2\rho})\right.\nonumber\\
            &+&\left.
    (\eta_{\alpha\sigma}p_{1\mu} -
          \eta_{\alpha\mu}p_{1\sigma})(\eta_{\beta\rho}
          p_{2\nu} - \eta_{\beta\nu}p_{2\rho})\right.\nonumber\\
          &+&\left.
    (\eta_{\alpha\rho}p_{1\nu} -
          \eta_{\alpha\nu}p_{1\rho})(\eta_{\beta\sigma}
            p_{2\mu} - \eta_{\beta\mu}p_{2\sigma})\right\}
\end{eqnarray}
The triple graviton vertex function is given by\cite{don}
\begin{eqnarray}
\tau^{\mu\nu}_{\alpha\beta,\gamma\delta}(k,q)&=&{i\kappa\over
2}\left\{ P_{\alpha\beta,\gamma\delta} \left[k^\mu k^\nu+(k-q)^\mu
(k-q)^\nu+q^\mu q^\nu-{3\over
2}\eta^{\mu\nu}q^2\right]\right.\nonumber\\
&+&\left.2q_\lambda
q_\sigma\left[I^{\lambda\sigma,}{}_{\alpha\beta}I^{\mu\nu,}
{}_{\gamma\delta}+I^{\lambda\sigma,}{}_{\gamma\delta}I^{\mu\nu,}
{}_{\alpha\beta}-I^{\lambda\mu,}{}_{\alpha\beta}I^{\sigma\nu,}
{}_{\gamma\delta}-I^{\sigma\nu,}{}_{\alpha\beta}I^{\lambda\mu,}
{}_{\gamma\delta}\right]\right.\nonumber\\
&+&\left.[q_\lambda
q^\mu(\eta_{\alpha\beta}I^{\lambda\nu,}{}_{\gamma\delta}
+\eta_{\gamma\delta}I^{\lambda\nu,}{}_{\alpha\beta})+ q_\lambda
q^\nu(\eta_{\alpha\beta}I^{\lambda\mu,}{}_{\gamma\delta}
+\eta_{\gamma\delta}I^{\lambda\mu,}{}_{\alpha\beta})\right.\nonumber\\
&-&\left.q^2(\eta_{\alpha\beta}I^{\mu\nu,}{}_{\gamma\delta}+\eta_{\gamma\delta}
I^{\mu\nu,}{}_{\alpha\beta})-\eta^{\mu\nu}q^\lambda
q^\sigma(\eta_{\alpha\beta}
I_{\gamma\delta,\lambda\sigma}+\eta_{\gamma\delta}
I_{\alpha\beta,\lambda\sigma})]\right.\nonumber\\
&+&\left.[2q^\lambda(I^{\sigma\nu,}{}_{\alpha\beta}
I_{\gamma\delta,\lambda\sigma}(k-q)^\mu
+I^{\sigma\mu,}{}_{\alpha\beta}I_{\gamma\delta,\lambda\sigma}(k-q)^\nu\right.\nonumber\\
&-&\left.I^{\sigma\nu,}{}_{\gamma\delta}I_{\alpha\beta,\lambda\sigma}k^\mu-
I^{\sigma\mu,}{}_{\gamma\delta}I_{\alpha\beta,\lambda\sigma}k^\nu)\right.\nonumber\\
&+&\left.q^2(I^{\sigma\mu,}{}_{\alpha\beta}I_{\gamma\delta,\sigma}{}^\nu+
I_{\alpha\beta,\sigma}{}^\nu
I^{\sigma\mu,}{}_{\gamma\delta})+\eta^{\mu\nu}q^\lambda q_\sigma
(I_{\alpha\beta,\lambda\rho}I^{\rho\sigma,}{}_{\gamma\delta}+
I_{\gamma\delta,\lambda\rho}I^{\rho\sigma,}{}_{\alpha\beta})]\right.\nonumber\\
&+&\left.[(k^2+(k-q)^2)\left(I^{\sigma\mu,}{}_{\alpha\beta}I_{\gamma\delta,\sigma}{}^\nu
+I^{\sigma\nu,}{}_{\alpha\beta}I_{\gamma\delta,\sigma}{}^\mu-{1\over
2}\eta^{\mu\nu}P_{\alpha\beta,\gamma\delta}\right)\right.\nonumber\\
&-&\left.(k^2\eta_{\gamma\delta}I^{\mu\nu,}{}_{\alpha\beta}+(k-q)^2\eta_{\alpha\beta}
I^{\mu\nu,}{}_{\gamma\delta})]\right\}
\end{eqnarray}
where we have defined
\begin{equation}
I_{\alpha\beta,\mu\nu}={1\over
2}(\eta_{\alpha\mu}\eta_{\beta\nu}+\eta_{\alpha\nu}\eta_{\beta\mu})
\end{equation}
and
\begin{equation}
P_{\alpha\beta,\mu\nu}=I_{\alpha\beta,\mu\nu}-{1\over
2}\eta_{\alpha\beta}\eta_{\mu\nu}
\end{equation}
The final ingredient which we need is the harmonic gauge graviton
propagator
\begin{equation}
D_{\alpha\beta,\mu\nu}(q)={i\over
q^2+i\epsilon}P_{\alpha\beta,\mu\nu}
\end{equation}

The leading component of the on-shell energy-momentum tensor
between charged vector meson states is then found, from Eq.
\ref{eqn:to}, to be
\begin{eqnarray}
<k_2,\epsilon_B|T_{\mu\nu}^{(0)}|k_1,\epsilon_A>&=&(k_{1\mu}k_{2\nu}+k_{1\nu}k_{2\mu})
\epsilon_B^*\cdot\epsilon_A\nonumber\\
&-&k_1\cdot\epsilon_B^*(k_{2\mu}\epsilon_{A\nu}+k_{2\nu}\epsilon_{A\mu}\nonumber\\
&-&k_2\cdot\epsilon_A(k_{1\nu}\eta_{B\mu}^*+k_{1\mu}\epsilon_{B\nu}^*)\nonumber\\
&+&(k_1\cdot
k_2-m^2)(\epsilon_{B\mu}^*\epsilon_{A\nu}+\epsilon_{B\nu}^*\epsilon_{A\mu})\nonumber\\
&-&\eta_{\mu\nu}[(k_1\cdot
k_2-m^2)\epsilon_B^*\cdot\epsilon_A-k_1\cdot\epsilon_B^*k_2\cdot\epsilon_A]\label{eqn:tm}
\end{eqnarray}
and the focus of our calculation is to evaluate the graviton loop
corrections to Eq. \ref{eqn:tm}, via the diagrams shown in Figure
1 and keeping only the leading nonanalytic terms, details of which
are described in the appendix.  Note that due to conservation of
the energy-momentum tensor---$\partial^\mu T_{\mu\nu}=0$---the
on-shell matrix element must satisfy the gauge invariance
condition
$$q^\nu<k_2,\epsilon_B|T_{\mu\nu}|k_1,\epsilon_A>=0$$
In our case, the leading order contribution satisfies this
condition
\begin{equation}
q^\mu<k_2,\epsilon_B|T_{\mu\nu}^{(0)}|k_1,\epsilon_A>=0
\end{equation}
and, in addition, the contributions of both diagrams 1a or 1b are
independently gauge-invariant
\begin{equation}
q^\mu Amp[a]_{\mu\nu}=q^\mu Amp[b]_{\mu\nu}=0
\end{equation}
and these strictures serve as an important check on our result.

Because of this gauge invariance condition, the results of these
calculations are most efficiently  expressed in terms of spin 1
form factors.  Indeed, due to covariance and gauge invariance the
form of the matrix element of $T_{\mu\nu}$ between on-shell spin 1
states must be expressible in the form
\begin{eqnarray}
&&<p_2,\epsilon_B|T_{\mu\nu}(x)|p_1,\epsilon_A>=-{e^{i(p_2-p_1)\cdot
x}\over \sqrt{4E_1E_2}}[2P_\mu P_\nu \epsilon_B^*\cdot
\epsilon_AF_1^{(S=1)}(q^2)\nonumber\\
&+&(q_\mu q_\nu-\eta_{\mu\nu}q^2)
\epsilon_B^*\cdot\epsilon_AF_2^{(S=1)}(q^2)\nonumber\\
&+&[P_\mu(\epsilon_{B\nu}^* \epsilon_A\cdot q-\epsilon_{A\nu}
\epsilon_B^*\cdot q)+P_\nu(\epsilon_{B\mu}^* \epsilon_A\cdot
q-\epsilon_{A\mu} \epsilon_B^*\cdot
q)]F_3^{(S=1)}(q^2)\nonumber\\
&+&\left[(\epsilon_{A\mu}
\epsilon_{B\nu}^*+\epsilon_{B\mu}^*\epsilon_{A\nu})q^2-(\epsilon_{B\mu}^*
q_\nu+\epsilon_{B\nu}^* q_\mu)\epsilon_A\cdot q\right.\nonumber\\
&+&\left.(\epsilon_{A\mu} q_\nu+\epsilon_{A\nu}
q_\mu)\epsilon_B^*\cdot q+2\eta_{\mu\nu}\epsilon_A\cdot q
\epsilon_B^*\cdot q\right]F_4^{(S=1)}(q^2)\nonumber\\
&+&{2\over m^2}P_\mu P_\nu \epsilon_A\cdot q \epsilon_B^*\cdot q
F_5^{(S=1)}(q^2)\nonumber\\
&+&{1\over m^2}(q_\mu q_\nu-\eta_{\mu\nu}q^2)\epsilon_B^*\cdot
q\epsilon_A\cdot qF_6^{(S=1)}(q^2)]
\end{eqnarray}
Using the feature that in the Breit frame for a nonrelativistic
particle the spin operator can be defined via
\begin{equation}
i(\hat{\epsilon}_B^*\times\hat{\epsilon}_A)_k=<1,m_f|S_k|1,m_i>
\end{equation}
we observe that
$F_1^{(S=1)}(q^2),F_2^{(S=1)}(q^2),F_3^{(S=1)}(q^2)$ correspond
exactly to their spin 1/2 counterparts while
$F_4^{(S=1)}(q^2),F_5^{(S=1)}(q^2),F_6^{(S=1)}(q^2)$ represent new
forms unique to spin 1.

In terms of these definitions, the tree level predictions can be
described as
\begin{eqnarray}
F_{1,tree}^{(S=1)}&=&F_{3,tree}^{(S=1)}=1\nonumber\\
F_{2,tree}^{(S=1)}&=&F_{4,tree}^{(S=1)}=-{1\over 2}\nonumber\\
F_{5,tree}^{(S=1)}&=&F_{6,tree}^{(S=1)}=0
\end{eqnarray}
while the results of the one loop calculation can be expressed as
\begin{itemize}
\item [a)] Seagull loop diagram (Figure 1a)
\begin{eqnarray}
F_{1,loop\, a}^{(S=1)}(q^2)&=&{GLq^2\over \pi}(0+3-1-{1\over
2})={3\over
2}{GLq^2\over \pi}\nonumber\\
F_{2,loop\, a}^{(S=1)}(q^2)&=&{GLm^2\over
\pi}(-5+2-2+4)=-{GLm^2\over
\pi}\nonumber\\
F_{3,loop\, a}^{(S=1)}(q^2)&=&{GLq^2\over \pi}(0+{3\over
2}-1-{1\over
2})=0\nonumber\\
F_{4,loop\, a}^{(S=1)}(q^2)&=&{GLm^2\over \pi}(0+1-1+{3\over 2})={3\over 2}{GLm^2\over \pi}\nonumber\\
F_{5,loop\, a}^{(S=1)}(q^2)&=&{GLm^2\over
\pi}(0-3+0+0)=-3{GLm^2\over
\pi}\nonumber\\
F_{6,loop\, a}^{(S=1)}(q^2)&=&{GLm^2\over \pi}(-5-{1\over
2}+0+3)=-{5\over 2}{GLm^2\over \pi}
\end{eqnarray}
\item [b)] Born loop diagram (Figure 1b)
\begin{eqnarray}
F_{1,loop\, b}^{(S=1)}(q^2)&=&{Gq^2\over \pi}[L({1\over
4}-3+2-{3\over
2})+S({1\over 16}-1+1+0)]={Gq^2\over \pi}({1\over 16}S-{9\over 4}L)\nonumber\\
F_{2,loop\, b}^{(S=1)}(q^2)&=&{Gm^2\over \pi}[S({7\over
8}-1+2-1)+L(1-3+4-3)]={Gm^2\over \pi}({7\over 8}S-L)\nonumber\\
F_{3,loop\, b}^{(S=1)}(q^2)&=&{Gq^2\over \pi}[S(0-{1\over
2}+{1\over 2}+{1\over 4})+L({1\over 6}-{5\over 4}+{3\over
4}+{7\over 12})]=
{Gq^2\over \pi}({1\over 4}S+{1\over 4}L)\nonumber\\
F_{4,loop\, b}^{(S=1)}(q^2)&=&{GLm^2\over \pi}(0-1+!-{3\over
2})+{Gq^2\over \pi}\left[L(-{17\over 8}+{3\over 8}-{1\over
2} +{7\over 8})\right.\nonumber\\
&+&\left.S(-{41\over 128}+{3\over 16}-{1\over 4}+{1\over
16})\right]=
-{3\over 2}{GLm^2\over \pi}-{Gq^2\over \pi}({11\over 8}L+{41\over 128}S))\nonumber\\
F_{5,loop\, b}^{(S=1)}(q^2)&=&{GLm^2\over \pi}(0+3+0+0)
+{Gq^2\over \pi}
\left[S({5\over 128}+{3\over 16}+0-{3\over 16})\right.\nonumber\\
&+&\left.L(0+{3\over 4}+0-{1\over 2})\right]=3{GLm^2\over
\pi}+{Gq^2\over \pi}({5\over 128}S+{1\over 4}L)\nonumber\\
F_{6,loop\, b}^{(S=1)}(q^2)&=&{Gm^2\over \pi}\left[S({43\over
64}-{1\over 8}+{1\over 4}-{1\over 8})+L({13\over 3}+{1\over
2}+{1\over 2} -{7\over 3})\right]\nonumber\\
&=&{Gm^2\over \pi}(3L+{43\over 64}S)
\end{eqnarray}
\end{itemize}
where we have divided each contribution into the piece which
arises from the first four bracketed pieces of the triple graviton
vertex above.\footnote{There exists no contribution to the
nonanalytic terms from the pieces in the fifth bracket since the
intermediate gravitons are required to be on-shell.}

The full results of this calculation can then be described via:
\begin{eqnarray}
F_1^{(S=1)}(q^2)&=&1+{Gq^2\over \pi}(-{3\over 4}L+{1\over 16}S)+\ldots\nonumber\\
F_2^{(S=1)}(q^2)&=&-{1\over 2}+{Gm^2\over \pi}(-2L+{7\over 8}S)+\ldots\nonumber\\
F_3^{(S=1)}(q^2)&=&1+{Gq^2\over \pi}({1\over 4}L+{1\over 4}S)+\ldots\nonumber\\
F_4^{(S=1)}(q^2)&=&-{1\over 2}+{Gq^2\over \pi}({11\over 8}L+{41\over 128}S)+\ldots\nonumber\\
F_5^{(S=1)}(q^2)&=&{Gq^2\over \pi}({1\over 4}L+{5\over
128}S)+\ldots\nonumber\\
F_6^{(S=1)}(q^2)&=&{Gm^2\over \pi}({1\over 4}L+{43\over
128}S)+\ldots
\end{eqnarray}
and we note that $F_{1,2,3,loop}^{(S=1)}(q^2)$ as found for unit
spin agree precisely with the forms $F_{1,2,3,loop}^{(S={1\over
2})}(q^2)$ determined previously for spin 1/2 and with
$F_{1,2,loop}^{(S=0)}(q^2)$ in the spinless case.  It is also
interesting that the loop contributions to the "new" form factors
$F_{4,loop}^{(S=1)}(q^2),F_{5,loop}^{(S=1)}(q^2)$ which have no
lower spin analog, vanish to order $q^0$ even though there exist
nonzero contributions from both loop diagrams individually.  Of
course, the nonrenormalization of $F_1^{(S=1)}(q^2=0)$ and
$F_3^{(S=1)}(q^2=0)$ required by energy-momentum and angular
momentum conservation is obtained, as required, meaning that, as
noted above, there exists no anomalous gravitomagnetic moment.

However, there is a new feature here that deserves notice. Working
in the Breit frame and assuming nonrelativistic motion, we have
the kinematic constraints
\begin{eqnarray}
\epsilon_A^0\simeq{1\over 2m}\hat{\epsilon}_A\cdot\vec{q},\qquad
\epsilon_B^0\simeq -{1\over
2m}\hat{\epsilon}_B^*\cdot\vec{q}\nonumber\\
\epsilon_B^*\cdot\epsilon_A\simeq
-\hat{\epsilon}_B^*\cdot\hat{\epsilon}_A-{1\over
2m^2}\hat{\epsilon}_B^*\cdot\vec{q}\hat{\epsilon}_A\cdot\vec{q}
\end{eqnarray}
we find that
\begin{eqnarray}
&&<p_2,\epsilon_B|T_{00}(0)|p_1,\epsilon_A>\simeq
m\left\{\hat{\epsilon}_B^*\cdot\hat{\epsilon}_AF_1^{(S=1)}(q^2)
+{1\over
2m^2}\hat{\epsilon}_B^*\cdot\vec{q}\hat{\epsilon}_A\cdot\vec{q}\right.\nonumber\\
&\times&\left.
[F_1^{(S=1)}(q^2)-F_2^{(S=1)}(q^2)-2(F_4^{(S=1)}(q^2)+F_5^{(S=1)}(q^2)
-{q^2\over 2 m^2}F_6^{(S=1)}(q^2))]
\right\}+\ldots\nonumber\\
&&<p_2,\epsilon_B|T_{0i}(0)|p_1,\epsilon_A>\simeq -{1\over
2}[(\hat{\epsilon}_B^*\times\hat{\epsilon}_A)\times\vec{q}]_iF_3^{(S-1)}(q^2)+\ldots
\end{eqnarray}
Then using the connections
\begin{eqnarray}
i\hat{\epsilon}_B^*\times\hat{\epsilon}_A&=&<1,m_f|\vec{S}|1,m_i>\nonumber\\
{1\over
2}(\epsilon_{Bi}^*\epsilon_{Aj}+\epsilon_{Ai}\epsilon_{Bj}^*)-{1\over
3}\delta_{ij}\hat{\epsilon}_B^*\cdot\hat{\epsilon}_A&=&<1,m_f|{1\over
2}(S_iS_j+S_jS_i)-{2\over 3}\delta_{ij}|1,m_i>\nonumber\\ \quad
\end{eqnarray}
between the Proca polarization vectors and the spin operator
$\vec{S}$ we can identify values for the gravitoelectric monopole,
gravitomagnetic dipole, and gravitoelectric quadrupole coupling
constants
\begin{eqnarray}
K_{E0}&=&mF_1^{(S=1)}(q^2=0)\nonumber\\
K_{M1}&=&{1\over
2}F_3^{(S=1)}(q^2=0)\nonumber\\
K_{E2}&=&{1\over
2m}\left[F_1^{(S=1)}(q^2=0)-F_3^{(S=1)}(q^2=0)-2F_4^{(S=1)}(q^2=0)-2F_5^{(S=1)}(q^2=0)\right]\nonumber\\
\end{eqnarray}
Taking $Q_g\equiv m$ as the gravitational "charge," we observe
that the tree level values---
\begin{equation}
K_{E0}=Q_g\qquad K_{M1}={Q_g\over 2m}\qquad K_{E2}={Q_g\over m^2}
\end{equation}
are {\it unrenormalized} by loop corrections.  That is to say, not
only does there not exist any anomalous gravitomagnetic moment, as
mentioned above, but also there is no anomalous gravitoelectric
quadrupole moment.

\section{Conclusion}

Above we have calculated the graviton loop corrections to the
energy-momentum tensor of a spin 1 system.  We have confirmed the
universality which was speculated in our previous work in that we
have verified that
\begin{eqnarray}
F_{1,loop}^{(S=0)}(q^2)&=&F_{1,loop}^{(S={1\over
2})}(q^2)=F_{1,loop}^{(S=1)}(q^2)\nonumber\\
F_{2,loop}^{(S=0)}(q^2)&=&F_{2,loop}^{(S={1\over
2})}(q^2)=F_{2,loop}^{(S=1)}(q^2)\nonumber\\
F_{3,loop}^{(S={1\over 2})}(q^2)&=&F_{3,loop}^{(S=1)}(q^2)
\end{eqnarray}
The universality in the case of the classical (square root)
nonanalyticities is not surprising and in fact is {\it required}
by the connection to the metric tensor.  In the case of the
quantum (logarithmic) nonanalyticities it is not clear why these
terms must be spin-independent. We also found additional form
factors for the spin 1 system and have shown that in addition to
the vanishing of the anomalous gravitomagnetic moment there cannot
exist any anomalous gravitoelectric quadrupole moment.  It is
tempting to conclude that the graviton loop correction
universality which we obtained holds for arbitrary spin. However,
it is probably not possible to show this by generalizing the
calculations above. Indeed the spin 1 result involves {\it
considerably} more computation than does its spin 1/2 counterpart,
which was already much more tedious than that for spin 0.  Perhaps
a generalization such as that used in nuclear beta decay can be
employed\cite{nbd}. Work is underway on such an extension and
results will be reported in an upcoming communication.

\begin{center}
{\bf Acknowledgement}
\end{center}

This work was supported in part by the National Science Foundation
under award PHY-02-42801.  Useful conversations with John Donoghue
and Andreas Ross are gratefully acknowledged, as is the
hospitality of Prof. A. Faessler and the theoretical physics group
from the University of T\"{u}bingen, where this paper was
finished.

\section{Appendix}

In this section we sketch how our results were obtained.  The
basic idea is to calculate the Feynman diagrams shown in Figure 1.
Thus for Figure 1a we find\cite{ppr}
\begin{equation}
Amp[a]_{\mu\nu}={1\over 2!}\int{d^4k\over
(2\pi)^4}{\epsilon_B^{*\beta}V^{(2)}_{\beta,\alpha,\lambda\kappa,\rho\sigma}(p_2,p_1)
\epsilon_A^\alpha
P[\alpha\beta;\lambda\kappa]P[\gamma\delta;\sigma\rho]
\tau_{\mu\nu}^{\alpha\beta,\gamma\delta}(k,q)\over
k^2(k-q)^2}\label{eqn:a}
\end{equation}
while for Figure 1b\cite{ppr}
\begin{eqnarray}
Amp[b]_{\mu\nu}&=&\int{d^4k\over (2\pi)^4}{1\over
k^2(k-q)^2((k-p)^2-m^2)}\nonumber\\
&\times&\epsilon_B^\beta
V^{(1)}_{\beta,\delta,\lambda\kappa}(p_2,p_1-k)\left(-\eta^{\delta\zeta}+{(p1-k)^\delta
(p_1-k)^\zeta\over
m^2}\right)\nonumber\\
&\times&V^{(1)}_{\zeta,\theta,\rho\sigma}(p1-k,p1)\epsilon_A^\theta
P[\alpha\beta;\lambda\kappa]P[\gamma\delta;\sigma\rho]
\tau_{\mu\nu}^{\alpha\beta,\gamma\delta}(k,q)\label{eqn:b}
\end{eqnarray}
Here the various vertex functions are listed in section 3, while
for the integrals, all that is needed is the leading nonanalytic
behavior.  Thus we use

\begin{eqnarray}
I(q)&=&\int{d^4k\over (2\pi)^4}{1\over k^2(k-q)^2}={-i\over
32\pi^2}(2L+\ldots)\nonumber\\
I_\mu(q)&=&\int{d^4k\over (2\pi)^4}{k_\mu\over k^2(k-q)^2}={i\over
32\pi^2}(q_\mu L+\ldots)\nonumber\\
I_{\mu\nu}(q)&=&\int{d^4k\over (2\pi)^4}{k_\mu k_\nu\over
k^2(k-q)^2}={-i\over 32\pi^2}(q_\mu q_\nu{2\over
3}L-q^2\eta_{\mu\nu}{1\over 6}L +\ldots)\nonumber\\
I_{\mu\nu\alpha}(q)&=&\int{d^4k\over (2\pi)^4}{k_\mu k_\nu
k_\alpha\over k^2(k-q)^2}={i\over 32\pi^2}(-q_\mu q_\nu q_\alpha
{L\over 2}\nonumber\\
&+&(\eta_{\mu\nu}q_\alpha+\eta_{\mu\alpha}q_\nu
+\eta_{\nu\alpha}q_\mu){1\over 12}Lq^2 +\ldots)\nonumber\\
\quad
\end{eqnarray}
for the "bubble" integrals and
\begin{eqnarray}
J(p,q)&=&\int{d^4k\over (2\pi)^4}{1\over
k^2(k-q)^2((k-p)^2-m^2)}={-i\over
32\pi^2m^2}(L+S)+\ldots\nonumber\\
J_\mu(p,q)&=&\int{d^4k\over (2\pi)^4}{k_\mu\over
k^2(k-q)^2((k-p)^2-m^2)}={i\over
32\pi^2m^2}\nonumber\\
&\times&[p_\mu((1+{1\over 2}{q^2\over m^2})L-{1\over 4}{q^2\over
m^2}S)-q_\mu(L+{1\over
2}S)+\ldots]\nonumber\\
J_{\mu\nu}(p,q)&=&\int{d^4k\over (2\pi)^4}{k_\mu k_\nu\over
k^2(k-q)^2((k-p)^2-m^2)}={i\over 32\pi^2m^2}\nonumber\\
&\times&[-q_\mu q_\nu(L+{3\over 8}S)-p_\mu p_\nu{q^2\over
m^2}({1\over 2}L+{1\over 8}S)\nonumber\\
&+&q^2\eta_{\mu\nu}({1\over 4}L+{1\over 8}S)+(q_\mu p_\nu+q_\nu
p_\mu)(({1\over 2}+{1\over 2}{q^2\over m^2})L+{3\over 16}{q^2\over
m^2 S})\nonumber\\
J_{\mu\nu\alpha}(p,q)&=&\displaystyle\int\frac{d^4k}{(2\pi)^4}
\frac{k_\mu k_\nu k_\alpha}{k^2(k-q)^2((k-p)^2-m^2)} \nonumber\\
&=& \frac{-i}{32\pi^2m^2}\bigg[ q_\mu q_\nu
q_\alpha\bigg(L+\frac5{16}S\bigg)+p_\mu p_\nu
p_\alpha\bigg(-\frac16 \frac{q^2}{m^2}\bigg) \nonumber\\
\nonumber&+&\big(q_\mu p_\nu p_\alpha + q_\nu p_\mu p_\alpha +
q_\alpha p_\mu p_\nu\big)\bigg(\frac13\frac{q^2}{m^2}L+
\frac1{16}\frac{q^2}{m^2}S\bigg)\nonumber\\&+&\big(q_\mu q_\nu
p_\alpha + q_\mu q_\alpha p_\nu + q_\nu q_\alpha p_\mu
\big)\bigg(\Big(-\frac13 - \frac12\frac{q^2}{m^2}\Big)L
-\frac{5}{32}\frac{q^2}{m^2}S\bigg)\nonumber\\
\nonumber &+&\big(\eta_{\mu\nu}p_\alpha + \eta_{\mu\alpha}p_\nu +
\eta_{\nu\alpha}p_\mu\big)\Big(\frac1{12}q^2L\Big)\nonumber\\
\nonumber&+&\big(\eta_{\mu\nu}q_\alpha + \eta_{\mu\alpha}q_\nu +
\eta_{\nu\alpha}q_\mu\big)\Big(-\frac16q^2L -\frac1{16}q^2S\Big)
\bigg]+\ldots\nonumber\\
\quad
\end{eqnarray}
for their "triangle" counterparts.  Similarly higher order forms
can be found, by either direct calculation or by requiring various
identities which must be satisfied when the integrals are
contracted with $p^\mu,q^\mu$ or with $\eta^{\mu\nu}$.  Using
these integral forms and substituting into Eqs. \ref{eqn:a} and
\ref{eqn:b}, one determines the results quoted in section 3.


\begin{thebibliography}{99}
\bibitem{fta} The Fourier transforms can be found in Appendix C of
\cite{gar}.
\bibitem{jkn} The $m^2$-independent piece of this matrix element
can be found from the simple energy-momentum tensor for the
electromagnetic field
$$T_{\mu\nu}^{em}=-F_{\mu\lambda}{F_\nu}^\lambda+
{1\over 4}\eta_{\mu\nu}F_{\lambda\delta}F^{\lambda\delta}.$$
\end{thebibliography}
\end{document}